\documentclass
[english,twocolumn,tightenlines,showpacs,amsmath,amssymb,pra]{revtex4}%
\usepackage{amsfonts}
\usepackage{amsmath}
\usepackage{amssymb}
\usepackage{graphicx}%
\begin{document}
\title{Coupling of a Single Diamond Nanocrystal to a Whispering-Gallery Microcavity:
Photon Transportation Benefitting from Rayleigh Scattering}
\author{Yong-Chun Liu}
\author{Yun-Feng Xiao\footnote{Author to whom correspondence should be addressed. Email: yfxiao@pku.edu.cn; URL: http://www.phy.pku.edu.cn/~yfxiao/index.html}}
\author{Bei-Bei Li}
\author{Xue-Feng Jiang}
\author{Yan Li}
\author{Qihuang Gong\footnote{Email: qhgong@pku.edu.cn}}
\affiliation{State Key Lab for Mesoscopic Physics, School of Physics, Peking University, P.
R. China.}

\begin{abstract}
We study the Rayleigh scattering induced by a diamond nanocrystal in a
whispering-gallery-microcavity--waveguide coupling system, and find that it
plays a significant role in the photon transportation. On one hand, this study
provides a new insight into future solid-state cavity quantum electrodynamics
toward strong coupling physics. On the other hand, benefitting from this
Rayleigh scattering, novel photon transportation such as dipole induced
transparency and strong photon antibunching can occur simultaneously. As
potential applications, this system can function as high-efficiency photon
turnstiles. In contrast to [B. Dayan \textit{et al.}, \textrm{Science}
\textbf{319},1062 (2008)], the photon turnstiles proposed here are highly
immune to nanocrystal's azimuthal position.

\end{abstract}

\pacs{78.35.+c, 42.50.Pq, 42.50.Ar}
\maketitle

Cavity quantum electrodynamics (CQED) studies light-matter interactions inside
a resonator, which offers an ideal platform for quantum optics
\cite{Science2002}. Tremendous progress has been made by coupling single
dipoles to different microcavities (for a review, see \cite{review}). Among
them, whispering-gallery mode (WGM)-type microcavity \cite{WGM} is most
promising due to its extremely high Q factor, small mode volume, excellent
scalability and ease for low-loss transport of nonclassical states using an
optical fiber \cite{thomas,disk}. On the other hand, nitrogen-vacancy (NV)
centers have recently emerged as an important candidate for quantum
information processing because they possess long-lived spin triplets at room
temperature \cite{gaebel,nv1}. Combining high-Q WGM microcavities and NV
centers represents a promising solid-state CQED system, and attracts much
attention very recently \cite{thomas,disk,PainterOE}.

One of the distinct properties of WGMs is that they belong to traveling waves,
unlike standing modes in a Fabre-Perot cavity. In other words, WGM
microcavities typically support twin modes, clockwise ($\mathrm{cw}$) and
counter-clockwise ($\mathrm{ccw}$) propagating waves with a degenerate
frequency. In this paper, we investigate the interaction of the twin WGMs
coupled to a NV center in a diamond nanocrystal. On one hand, we find that not
only the dipole transition of the NV center but also the Rayleigh scattering
by the nanocrystal itself play significant roles in the coupled system. Thus
strong coupling condition in such CQED system should be re-defined. On the
other hand, nonclassical effects are predicted benefitting from the
nanocrystal scattering. Dipole induced transparency (DIT) \cite{DITPRL2006}
appears in the presence of the strong Rayleigh scattering, accompanying strong
photon antibunching in transmitted fields. Moreover, high-efficiency photon
turnstiles can be implemented highly immune to the azimuthal position of the nanocrystal.

\begin{figure}[tb]
\begin{center}
\centerline{\includegraphics[width=7.5cm]{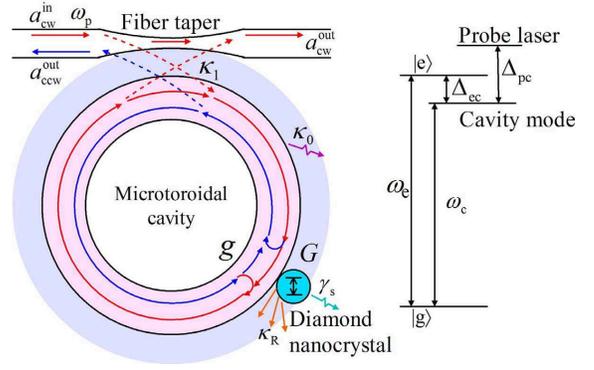}}
\end{center}
\caption{Schematic illustration of the nanocrystal-microcavity-waveguide
coupling system.}%
\label{fig1}%
\end{figure}

Figure \ref{fig1} describes the present coupling system. A diamond nanocrystal
embodying a single NV center is located on the surface of a microtoroidal
cavity \cite{vahala}. A tapered fiber is used to couple light with frequency
$\omega_{p}$ in and out of the microcavity. Under the rotating wave
approximation, the Hamiltonian of the coupled system can be written as
($\hbar=1$)
\begin{align}
H  &  =H_{0}+H_{1}+H_{2},\nonumber\\
H_{0}  &  =\omega_{e}\left\vert e\right\rangle \left\langle e\right\vert
+\sum\limits_{m}{\omega_{c}a_{m}^{\dag}a_{m}}+\sum\limits_{k=1}^{3}%
\sum\limits_{j}{\omega_{kj}b_{kj}^{\dag}b_{kj},}\nonumber\\
H_{1}  &  =\sum\limits_{m}{{G_{m}a_{m}^{\dag}\sigma_{-}}}+\sum\limits_{j}%
{{G_{j}b_{1j}^{\dag}\sigma_{-}+H.c.},}\nonumber\\
H_{2}  &  =\sum\limits_{m,m^{\prime}}{g_{m,m^{\prime}}a_{m}^{\dag}%
a_{m^{\prime}}}+{\sum\limits_{j}}({{{g_{\text{\textrm{cw}},j}%
a_{\text{\textrm{cw}}}^{\dag}b_{2j}+H.c.)}}}\nonumber\\
&  {{+{\sum\limits_{j}}({{{g_{\text{\textrm{ccw}},j}a_{\text{\textrm{ccw}}%
}^{\dag}b_{3j}+H.c.)}}}.}}%
\end{align}
Here summation indices ($m$, $m^{\prime}$) run through \textrm{cw},
\textrm{ccw}; $a_{m}$ and $b_{kj}$ $(k=1,2,3)$ denote annihilation operators
of the cavity and the reservoir modes, respectively; $\sigma_{-}=\left\vert
g\right\rangle \left\langle e\right\vert $ and $\sigma_{+}=\left\vert
e\right\rangle \left\langle g\right\vert $ stand for the descending and
ascending operators of the dipole transition at zero-phonon line. The
Hamiltonian $H_{0}$ describes the free evolution of the system consisting of
the NV center, WGMs and reservoir modes, with frequency $\omega_{e}$,
$\omega_{c}$, $\omega_{kj}$, respectively. The first (second) term in $H_{1}$
characterizes the dipole interactions between the NV center and WGMs
(reservoir modes), with the coupling strength $G_{m}$ ($G_{j}$). The first
term in $H_{2}$ describes the nanocrystal-induced scattering into the same
($m=m^{\prime}$) or the counter-propagating ($m\neq m^{\prime}$) WGM fields
with strengths $g_{m,m^{\prime}}$, while the second term represents the
WGMs-reservoir scattering coefficients $g_{m,j}$.

In the full quantum theory, the coherent coupling strengths $G_{m}$ and
$G_{j}$ can be calculated by $G_{m}\equiv G=\mu({\omega_{c}/(2\hbar
\varepsilon_{0}\varepsilon_{s}V_{c}))}^{1/2}f_{c}\left(  \overrightarrow
{r}\right)  $ and $G_{j}=\mu({\omega_{1j}/(2\hbar\varepsilon_{0}%
\varepsilon_{s}V_{1j}))}^{1/2}$, where $\mu=2.74\times10^{-29}$ \textrm{C}%
$\mathrm{\cdot}$\textrm{m} represents the dipole moment of the NV center
transition. ${\varepsilon_{0}}$ is the electric permittivity of the vacuum and
${\varepsilon}_{s}=1$ denotes the relative permittivity of the surrounding
medium (vacuum here). ${V_{c}}$ and ${V_{1j}}$ are the quantized volumes of
the cavity and the reservoir modes interacting with the dipole, respectively.
$f_{c}\left(  \vec{r}\right)  =|E(\vec{r})/E_{\max}|$ is the normalized field
distribution function of WGMs. For the subwavelength nanocrystal, the above
scattering interaction can be modeled in a dipole approximation where the
electric field of the input wave (either WGMs or reservoir modes) induces a
dipole moment in the scatterer, namely, \textit{scattering induced dipole}. In
the case of elastic Rayleigh scattering, the coupling strengths are
$g_{m,m^{\prime}}\equiv-g=-\alpha f_{c}^{2}(\vec{r}){\omega_{c}/(2{V_{c}})}$
and $g_{m,j}\equiv-g_{R}=-\alpha f_{c}(\vec{r}){\omega_{c}}(\hat{n}_{m}%
\cdot\hat{n}_{k,j})/(4V_{c}V_{k,j})^{1/2}$ ($m=\mathrm{cw},k=2$ or
$m=\mathrm{cw},k=2$)\cite{mazzei}. Here $\alpha=4\pi R^{3}(\varepsilon
_{d}-\varepsilon_{s})/(\varepsilon_{d}+2\varepsilon_{s})$ is the
polarizability for the spherical nanocrystal with radius $R$ where
$\varepsilon_{d}=2.4^{2}$ denotes the electric permittivity of the diamond.
$\hat{n}_{m}$ and $\hat{n}_{k,j}$ are unit vectors of the fields.

We can find that both the NV center-reservoir and WGM-reservoir coupling
interactions have the same format. By using Weisskopf-Wigner approximation,
the coupling of the NV center to the reservoir can be regarded as a decay of
the excited state, i.e., the well-known spontaneous emission with rate
$\gamma_{s}$; while the scattering of WGMs to the reservoir can be modeled by
an energy damping of WGMs, with the damping rate $\kappa_{R}={\alpha^{2}%
f_{c}^{2}(\vec{r})\varepsilon_{s}^{3/2}\omega_{c}^{4}/(6\pi c^{3}V_{c})}$
\cite{mazzei}, where $c$ is the speed of light in vacuum. Transforming
travelling to standing modes, $a_{\pm}=(a_{\mathrm{cw}}\pm a_{\mathrm{ccw}%
})/\sqrt{2}$, the equations of motion for the coupled system are given by
\cite{Gardiner}
\begin{align}
\frac{da_{+}}{dt}  &  =(2ig-\kappa_{+})a_{+}-\sqrt{2}iG\sigma_{-}-\sqrt
{\frac{\kappa_{1}}{2}}a_{\mathrm{cw}}^{\mathrm{{in}}}+\mathbf{\hat{f}}%
_{+},\label{eq:motion_ap}\\
\frac{da_{-}}{dt}  &  =-\kappa_{-}a_{-}-\sqrt{\frac{\kappa_{1}}{2}%
}a_{\mathrm{cw}}^{\mathrm{{in}}}+\mathbf{\hat{f}}_{-},\\
\frac{d\sigma_{-}}{dt}  &  =-(i\Delta_{\mathrm{ec}}+\frac{\gamma_{s}}%
{2})\sigma_{-}+i\sqrt{2}Ga_{+}{\sigma_{z}}+\mathbf{\hat{f}}_{1}. \label{eq:sm}%
\end{align}
Here ${\sigma_{z}}\equiv\left\vert e\right\rangle \left\langle e\right\vert
-\left\vert g\right\rangle \left\langle g\right\vert $, $\Delta_{\mathrm{ec}%
}\equiv\omega_{e}-{\omega_{c}}$, $\kappa_{+}=\kappa_{R}+({\kappa_{0}%
+\kappa_{1})/2}$, $\kappa_{-}=({\kappa_{0}+\kappa_{1})/2}$; ${\kappa
_{0}=\omega_{c}/Q}_{0}$ denotes the intrinsic damping of the WGMs with
${Q}_{0}$ being the intrinsic quality factor; ${\kappa_{1}}$ stands for the
cavity-taper coupling strength; $a_{\mathrm{cw}}^{\mathrm{in}}$ is the input
field. The operators $\mathbf{\hat{f}}_{+}$, $\mathbf{\hat{f}}_{-}$,
$\mathbf{\hat{f}}_{1}$ are the noise operators that conserve the commutation
relations at all times. One can find that the symmetric mode $a_{+}$ is
strongly coupled to the NV center (through dipole interaction) and the
nanocrystal (through the scattering), while the anti-symmetric mode $a_{-}$
keeps uncoupled to them.

When the cavity is excited by a weak monochromatic field (i.e., weak-field
approximation), the NV center is predominantly in the ground state. Thus
$\sigma_{z}$ can be substituted for its average value of $-1$, and Eq.
(\ref{eq:sm}) becomes linear. Utilizing the standard input-output formalism
$a_{m}^{\mathrm{{out}}}=a_{m}^{\mathrm{in}}+\sqrt{\kappa_{1}}a_{m}$
\cite{Gardiner}, we obtain the outputs
\begin{align}
a_{\mathrm{cw}}^{\mathrm{{out}}}  &  =\left[  1+\frac{\kappa_{1}}{2}\left(
\frac{1}{D_{+}}+\frac{1}{D_{-}}\right)  \right]  a_{\mathrm{cw}}%
^{\mathrm{{in}}}+\mathbf{\hat{f}}_{2}^{\prime},\label{aoutcw}\\
a_{\mathrm{ccw}}^{\mathrm{{out}}}  &  =\frac{\kappa_{1}}{2}\left(  \frac
{1}{D_{+}}-\frac{1}{D_{-}}\right)  a_{\mathrm{cw}}^{\mathrm{{in}}%
}+\mathbf{\hat{f}}_{3}^{\prime}, \label{aoutccw}%
\end{align}
where $D_{+}=i(\Delta_{\mathrm{pc}}+2g)-\kappa_{+}+2G^{2}/[i(\Delta
_{\mathrm{pc}}-\Delta_{\mathrm{ec}})-\gamma_{s}/2]$, $D_{-}=i\Delta
_{\mathrm{pc}}-\kappa_{-}$, with $\Delta_{\mathrm{pc}}\equiv\omega_{p}%
-\omega_{c}$, $\mathbf{\hat{f}}_{2}^{\prime}$ and $\mathbf{\hat{f}}%
_{3}^{\prime}$ are noise operators. We take the cold reservoir limit where the
reservoir modes are all initially in vacuum states. Thus the expectation
values of all noise operators can be neglected because they are annihilated
when acting on the initial vacuum states. Then the transmission $\left\langle
a_{\mathrm{cw}}^{\mathrm{{out}}\dag}a_{\mathrm{cw}}^{\mathrm{{out}}%
}\right\rangle /\left\langle a_{\mathrm{cw}}^{\mathrm{{in}}\dag}%
a_{\mathrm{cw}}^{\mathrm{{in}}}\right\rangle $ and reflection $\left\langle
a_{\mathrm{ccw}}^{\mathrm{{out}}\dag}a_{\mathrm{ccw}}^{\mathrm{{out}}%
}\right\rangle /\left\langle a_{\mathrm{cw}}^{\mathrm{{in}}\dag}%
a_{\mathrm{cw}}^{\mathrm{{in}}}\right\rangle $ can be determined.

\begin{figure}[ptb]
\begin{center}
\includegraphics[width=7.5cm,angle=0]{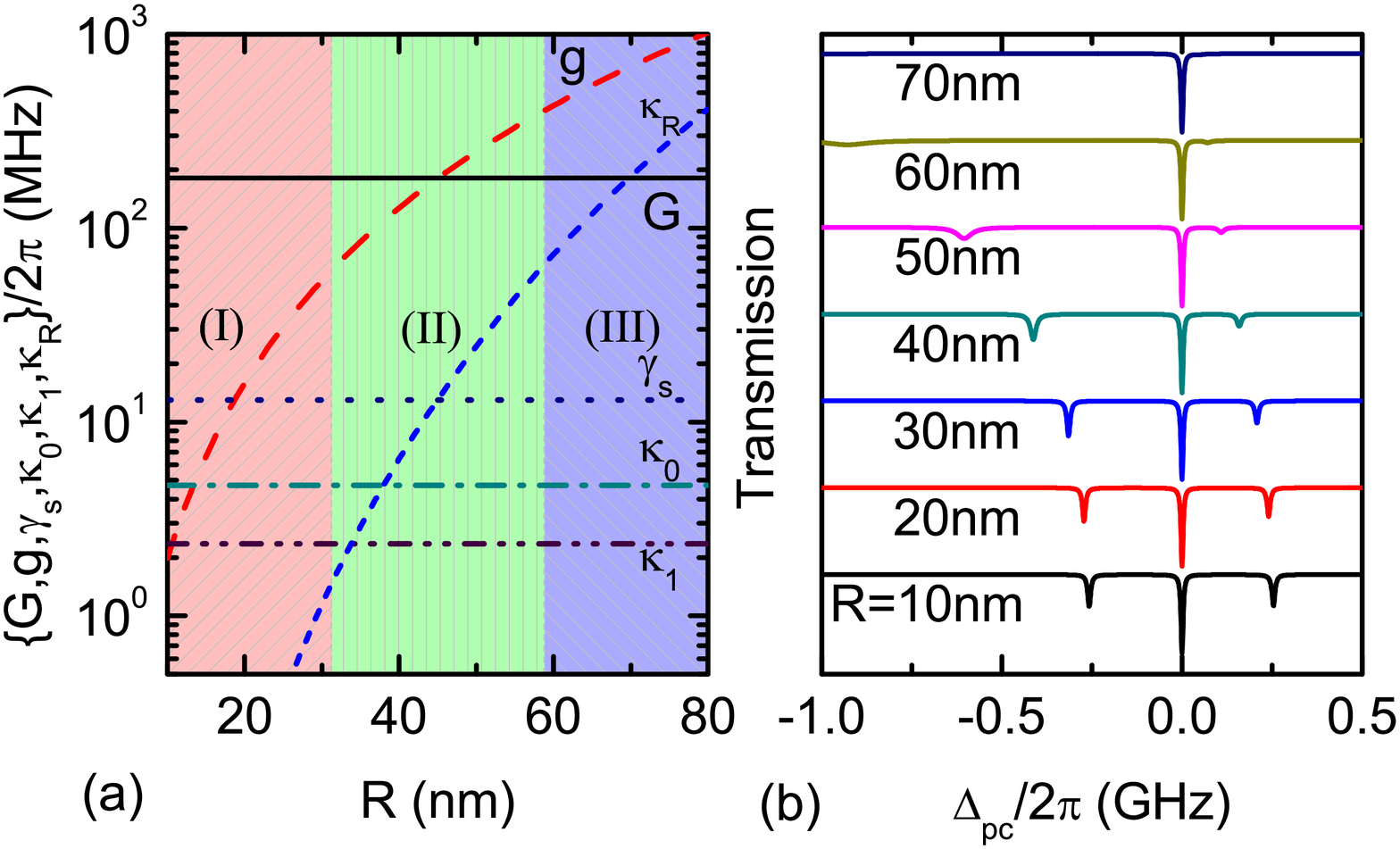}
\end{center}
\caption{(a) Parameters \{$G$, $g$, $\gamma_{s}$, $\kappa_{0}$, $\kappa_{1}$,
$\kappa_{R}$\} vs. the radius $R$\ of the nanocrystal. (b) The transmission
spectra for various $R$. Here ${Q}_{0}${ }${=10}^{8}$, $\kappa_{1}%
=0.5\kappa_{0}$ and $\Delta_{ec}=0.$}%
\label{fig2}%
\end{figure}

Under practical parameters $\gamma_{s}/2\pi=13$ \textrm{MHz}, ${Q}_{0}${
}${=10}^{8}$, $\kappa_{1}=0.5\kappa_{0}$, $V_{c}\sim200$ $\mathrm{\mu m}^{3}$,
$f_{c}\left(  \overrightarrow{r}\right)  \sim0.47$, we obtain $\left\{
G,\kappa_{0},\kappa_{1}\right\}  /2\pi\sim$\ $\left\{  180,4.7,2.35\right\}
$\textrm{MHz}. We will use these parameters unless otherwise specified. Figure
\ref{fig2}(a) plots these parameters depending on the particle radius $R$. We
can approximately divide Fig. \ref{fig2}(a) into three regions. In region I,
$G$ far exceeds any other parameters for a small diamond nanocrystal. The NV
center couples to the twin WGMs, creating a pair of standing modes. There
exist three separate resonance dips in transmission spectrum: one at the
original zero detuning due to the decoupling between the anti-symmetric mode
and the NV center, and the other two splitted by $2\sqrt{2}G$ due to the
strong coupling between the symmetric standing mode and the NV center. The
linewidth of the central dip is $\kappa_{0}+\kappa_{1}$; while the linewidths
of the two sideward dips are $\kappa_{R}+(\kappa_{0}+\kappa_{1}+\gamma_{s}%
)/2$, which can be interpreted by the dressed state generated by the strong coupling.

With the increase of $R$, the coefficient $g$ grows rapidly. In region II, $g$
becomes comparable with $G$, while $\kappa_{R}$ is still much smaller than $G$
because $g(\kappa_{R})\propto R^{3}(R^{6})$. The scattering not only moves the
two sideward resonances with detunings $-g\pm(g^{2}+2G^{2})^{1/2}$, but also
increases their linewidths and decreases the coupling efficiencies. When the
size of the nanocrystal is large enough (region III), the scattering damping
rate $\kappa_{R}$ becomes comparable to or even exceeds $G$. In this
situation, the scattering damping rate is so large that the two sideward dips
nearly vanish. In all cases, the central dip keeps unchanged, and its coupling
depends mainly on $\kappa_{0}$ and $\kappa_{1}$.

Transmission spectra for different $R$ varying from $10$ to $70$ nm are ploted
in Fig. \ref{fig2}(b), confirming the above analysis. As the
nanocrystal-induced scattering does occur, the new strong coupling condition
requires that $G\gg\{{\kappa_{+},\gamma_{s}\}}$, corresponding to regions I
and II. This provides new insights into future solid-state CQED.

In the region where the present strong coupling condition does not match,
interesting DIT effects in the transmission spectrum can be predicted. Let us
consider the conditions for a general DIT phenomenon. First, in the absence of
the dipole, the new critical coupling condition ($\kappa_{1}^{2}=\kappa
_{0}^{2}+4g^{2}$) is required to create a relatively broad platform with a low
transmission. This can be implemented though the destructive interference
among the directly transmitted field uncoupled to the microcavity and the
output field via cavity modes, as discussed in detail in the following.
Second, the linewidth of the resonance should be larger than the dipole-cavity
coupling strength ($\kappa_{+},\kappa_{-}>G$, \textit{i.e.}, the bad cavity
limit). These conditions are verified in Fig. 3, which presents the
transmission, reflection and energy loss in different conditions.

In the weak scattering case ($g\ll\kappa_{0}$), the initial critical coupling
condition without dipole is simplified as $\kappa_{1}=\kappa_{0}$. For
cavities with high intrinsic quality factor, the mode splitting occurs instead
of DIT, as shown in Figs. 3(a)-3(b). This is because the dipole-cavity
coupling strength exceeds the resonance linewidth. While a cavity with low
intrinsic quality factor can produce DIT as shown in Figs. 3(c)-3(d), but
there are significant energy losses, yielding the low DIT peak, as well as low
on/off contrast ratio.

In the strong scattering case ($g\gg\kappa_{0}$), however, we find that the
nanocrystal-induced Rayleigh scattering plays a constructive role in
generating DIT and can significantly enhance it, as demonstrated in Figs.
3(e)-3(h). In general, the field radiated by the NV center dipole takes part
in the interference mentioned above. If we control the detuning\textit{
}$\Delta_{\mathrm{ec}}=-g-G^{2}/g$\textit{ }so that the radiated field
intensity has a maximum value at $\Delta_{\mathrm{pc}}\simeq-g$ where the
initial critical-coupling point is located at. The DIT window (also at
$\Delta_{\mathrm{pc}}\simeq-g$) with a high efficiency (the peak value exceeds
$0.8$) appears, because the intrinsic loss is far smaller than the
cavity-taper coupling rate ($\kappa_{1}\sim2g\gg\kappa_{0}$).

\begin{figure}[ptb]
\begin{center}
\includegraphics[width=7.5cm,angle=0]{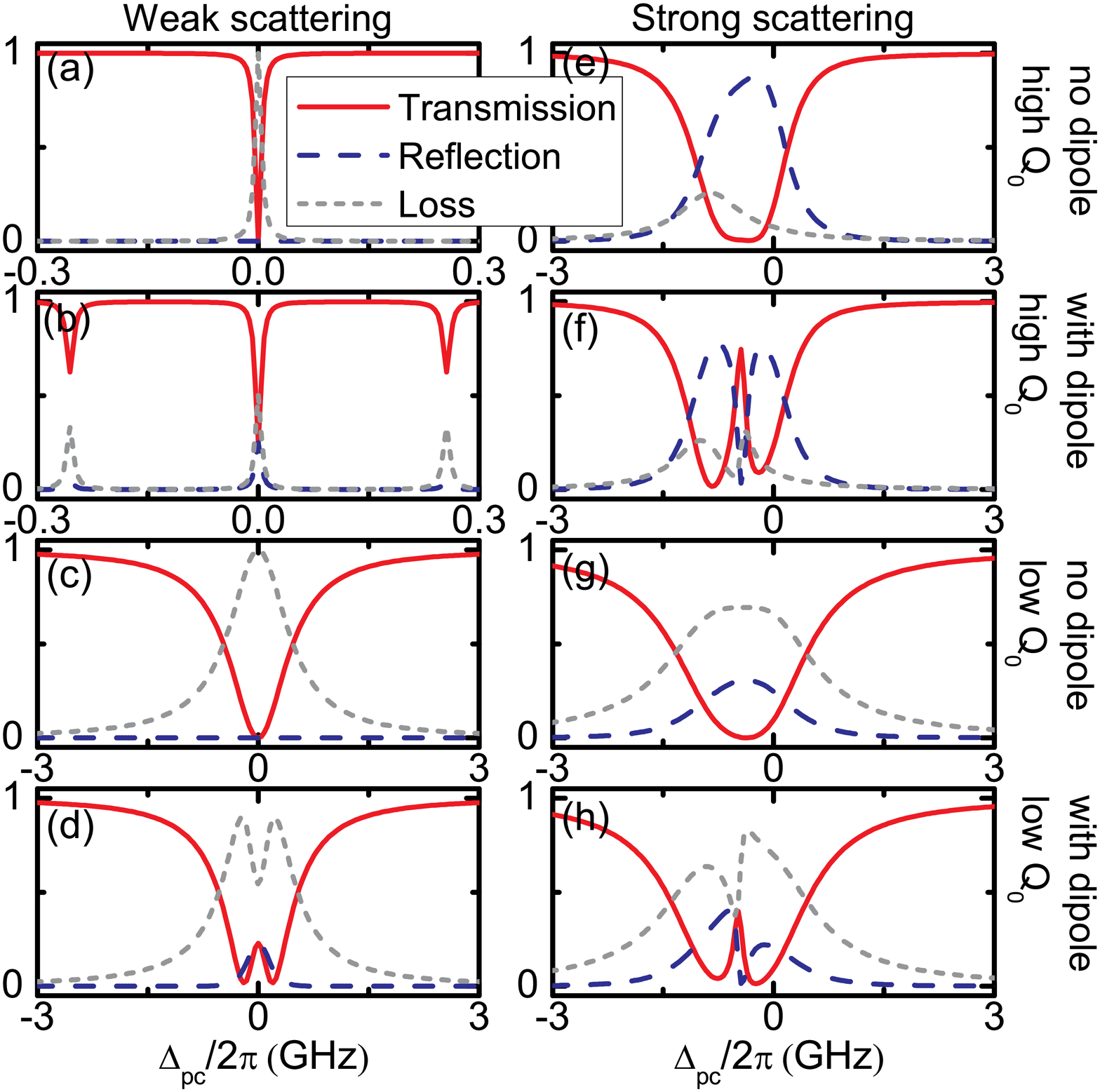}
\end{center}
\caption{Transmission, reflection and energy loss of the present system
working on the critical coupling point $\kappa_{1}=(\kappa_{0}^{2}%
+4g^{2})^{1/2}$. (a)-(d), weak scattering case with $R=6$ nm and
$\Delta_{\mathrm{ec}}=0$. (e)-(h), strong scattering case with $R=60$ nm and
$\Delta_{\mathrm{ec}}=-g-G^{2}/g$. The high and low $Q_{0}$ are $10^{8}$ and
$10^{6}$, respectively.}%
\label{fig3}%
\end{figure}

\begin{figure}[tb]
\begin{center}
\includegraphics[width=7.5cm,angle=0]{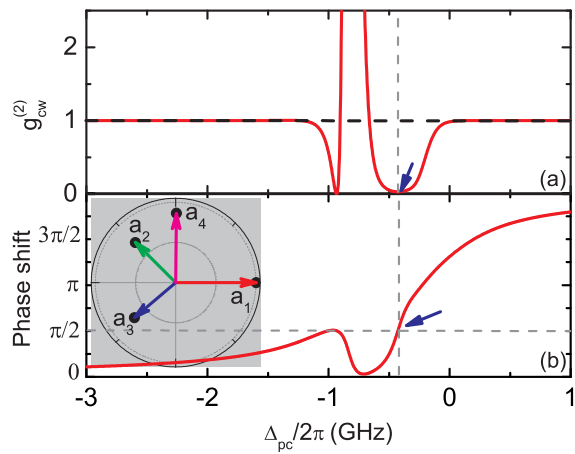}
\end{center}
\caption{Normalized second-order autocorrelation functions $g_{\mathrm{cw}%
}^{(2)}$ (a) and phase shift (b) of the corresponding transmitted field in
Fig. 3(f). The dashed vertical lines and blue arrows indicate the position of
$\Delta_{\mathrm{pc}}{=-g}$. Inset: Four components of the transmitted field
in complex plane (see text).}%
\label{fig4}%
\end{figure}

High-efficiency DIT holds great potential for photon turnstiles
\cite{Dayan,painterPRA}. We are thus interested in the photon statistics of
the output \cite{CarmichaelBook,Carmichael1985}, described by the second-order
correlation functions $g_{\mathrm{m}}^{(2)}=\left\langle (a_{\mathrm{m}%
}^{\mathrm{{out}}\dag})^{2}(a_{\mathrm{m}}^{\mathrm{{out}}})^{2}\right\rangle
/\left\langle a_{\mathrm{m}}^{\mathrm{{out}}\dag}a_{\mathrm{m}}^{\mathrm{{out}%
}}\right\rangle ^{2}$. Figure \ref{fig4}(a) depicts the strong photon
antibunching of the transmitted field with a large bandwidth, corresponding to
the case in Fig. \ref{fig3}(f). This is because the net transmitted field
mainly originates from the NV center dipole and a single dipole cannot emit
two or more photons simultaneously. Actually, excitation of the system by a
first photon blocks the transmission of a second photon, known as photon
blockade \cite{PhotonBlockade1997}. It is of importance that the strong
antibunching ($g_{\mathrm{cw}}^{(2)}=0$) occurs exactly at the DIT window,
which guarantees the high efficiency of the antibunched light.

In the DIT window, the phase shift $\phi_{\mathrm{cw}}$ of the transmitted
field (defined as $\phi_{\mathrm{cw}}\equiv\arg[a_{\mathrm{cw}}^{\mathrm{{out}%
}}/a_{\mathrm{cw}}^{\mathrm{in}}]$) has also been plotted in Fig.
\ref{fig4}(b), which exhibits a very large dispersion in the region near
$\Delta_{\mathrm{pc}}=-g$. This enables us to control the phase shift and
group delay of the output field. It is interesting that the transmitted field
experiences a $\pi/2$ phase shift at the DIT peak, different from the previous
DIT in \cite{DITPRL2006} (zero shift). To explicitly explain this, from Eq.
(\ref{aoutcw}), we note that the transmitted field is an interference of four
components: directly transmitted field uncoupled to the microcavity, output
field via the anti-symmetric mode, output field via the Rayleigh scattering,
and output field via the NV-center dipole interaction, denoted by
$a_{k=1,2,3,4}$, respectively. In the inset of Fig. \ref{fig4}(b) we plot the
four components for $\Delta_{\mathrm{pc}}=-g$\ in the complex plane. Here, we
find $a_{1}=a_{\mathrm{cw}}^{\mathrm{in}}$ and $a_{2}=(\kappa_{1}%
/2)a_{\mathrm{cw}}^{\mathrm{in}}/(i\Delta_{\mathrm{pc}}-\kappa_{-})$, which
cannot be tuned via the nanocrystal; $a_{3}=(\kappa_{1}/2)a_{\mathrm{cw}%
}^{\mathrm{in}}/(i(\Delta_{\mathrm{pc}}+2g)-\kappa_{+})$ depends on the
scattering strength $g$ while $a_{4}$ depends mainly on $G$ and $\Delta
_{\mathrm{ec}}$. It can be seen that $a_{1},a_{2},a_{3}$ destructively
interference, creating zero net field. $a_{4}$ has a large amplitude with the
relative phase $\pi/2$. This further explains the strong antibunching of the
transmitted field in the DIT window.

Finally, in Ref. \cite{Dayan}, photon turnstiles are realized by coupling
microcavity to single cooled caesium atoms. Standing wave modes built there
are induced by scattering from defects and surface roughness of the
microtoroidal cavity, and thus photon turnstiles are sensitive to the atom's
azimuthal position. Averaging over the azimuthal angle results in a
significant reduction of the on/off contrast ratio of transmission in the DIT
window (see Fig. 1(E) in Ref. \cite{Dayan}). While in the present paper,
Rayleigh scattering is induced by the diamond nanocrystal itself. Thus these
devices are highly immune to azimuthal position of the nanocrystal. This is of
special importance when multi-nanocrystal are involved in the microcavity
system. For example, the dynamics of the coupling system will strongly depend
on the inter-positions of these nanocrystals.

In summary, we have analyzed a single diamond nanocrystal with a NV center
coupled to high-Q counter-propagating twin WGMs of a microtoroidal cavity. It
is found that the Rayleigh scattering induced by the nanocrystal itself not
only induces strong interaction between the twin WGMs, but also results in the
scattering loss of the WGMs. This provides a new insight into future
solid-state CQED. We also reveal that Rayleigh scattering can play a positive
role in obtaining DIT with high efficiency, accompanying strong photon
antibunching. The system can perform as photon turnstiles which are highly
immune to the azimuthal position of nanocrystal. Our investigation can be of
practical use in controlled interactions of single quanta and scalable quantum logic.

\textit{Note added - We note a recent publication \cite{scattering}\ after the
submission of our work.}

\begin{acknowledgments}
This work was supportted by NSFC (Nos. 10821062 and 11004003) and 973 program
(No. 2007CB307001).  Yun-Feng Xiao was also supported by the Research Fund for the Doctoral Program of Higher Education (Grant No. 20090001120004)
and the Scientific Research Foundation for the Returned Overseas Chinese Scholars.
\end{acknowledgments}

\end{document}